# The hands of time: Moving my body to keep time order in the brain

Julien Lagarde

Université Pau et pays de l'Adour, laboratory MEPS – STAPS Tarbes

**Email: Julien.lagarde@univ-pau.fr**

**Acknowledgements**

This study was financially supported by the European Project ENTIMEMENT, H2020-FETPROACT-2018, Grant Number 824160. The author wishes to thank Anders Ledberg, Sofiane Ramdani, Nicolas Bouisset, Julia Ayache, Viktor Jirsa, Mark Herzlich, Franck Jourdan, and Daniel Bennequin for insightful discussions and wonderful scholar experiences. To my beloved father.

**Preamble**

"Now is the winter of our discontent, made glorious summer by this sun of York" (WS, Richard III)

"All those moments will be lost in time, like tears in the rain" (Roy Batty, Blade Runner)

**Abstract**

The brain is very often viewed as a network, be it at small scale made of cells, mostly neurons, or at larger scale made of neuronal assemblies. Here we introduce a conjecture, in the spirit of a philosophical though experiment, which proposes that the present cannot be obtained from within such networks, and that this limitation imposes burdens on network efficiency in information processing. We aim to argue this conjecture imposes recurrent contacts from within the brain to outside in the physical world via behaviour, which create a flow of time stamps. This though experiment may contribute to make the divide between the foci toward inside versus outside, for example opposing ecological psychology and many frameworks adopted in neurosciences, superfluous. This piece proposes an ambulation triggered by a thought experiment: What if I was a neuron listening to another one and talking to a third? It is a modest attempt to walk in the footsteps of classical thought experiments, like Molyneux's problem, the imitation game and the anti-sequel Chinese room, key gedankenexperiments in an elevator in physics, or the cogito in philosophy.

**Overview**

This work introduces a theoretical conjecture based on a thought experiment, in order to contribute to "darn" the divide between two theoretical inclinations: Ecological psychology, rooted in the relation between the animal and the environment, and neuroscience, which might at times lean toward isolating the inner activity of the brain from the relation to the environment. The former is tempted by a sort of neuronal eliminativism, and formulates criticism regarding the latter, for calling systematically recourse to neuronal explanations to most of the problems the animal faces. The latter may too often forget anchoring the brain to behavioral functions. To a very large extent, contemporaneous approaches in neuroscience consider brains as interconnected distributed systems, represented by graphs of neurons and synaptic connections, spanning a wide range of spatial scales. This is currently extended to higher order interactions, where hypergraph or simplicial generalize graph. Crucially, it is well known that the transmission of action potentials between distant areas in the brain are delayed. Using a very simplified model of network with such delayed connections, we ask how to establish a temporal order between the activities of the nodes in the network, strictly from within the network isolated from the environment. This tracking of the order of activity would support a constant record of which variation of activity is past or coincident to which other activity. Simply put, it would enable to function, that is, subserving cognition, perception, decision, action, attention, or memory, according to the rule "a is before b", which affords a basic ordering of past, present, and future activities. We propose the expression t-present to summarize this creation and maintenance of time order and flow.

From that starting point we propose that to get this t-present, delayed communication in noisy networked brain imposes physical interaction with the environment. Based on the analysis of a toy model consisting in three neurons arranged on a ring, we suggest, using a though experiment, that an updated t-present cannot be obtained from inside the networks, and that by contrast it could be easily obtained by direct contact, that is an interaction, with the surrounding physical environment. To get such contact autonomously, any movement of the body may be sufficient, which would include the motion of sensory systems in the spatial external frame. Phasic or continuous interaction may provide transient timestamps, obtained locally in sensorimotor brain networks, and which potentially propagate to other parts of the brain, underpinning specific functions not necessarily restricted to temporal processing, via a mediating readout network or maybe simply by a direct coupling. Incidentally, this research program could provide a linkage between biological scales far apart a priori: Bodily actions in the environment and neural networks dynamics.

## 1) Introduction

In order to nourish theoretical, perhaps philosophical, thinking in neuroscience and cross-disciplinary fertilization, we introduce the divide between ecological psychology, notably of direct perception (Gibson, 1966), and network neurosciences (Basset, Sporns, 2017; Bullmore and Sporns, 2009). These traditions have largely evolved in parallel. Here we ask whether a fundamental problem in network neuroscience—the ordering of events in time—may help reconnect them. We propose a conjecture, introduced by a though experiment, which may help bind the two approaches. We hasten to assert that this work is speculative, and it belongs to the type of contribution nowadays referred to as an opinion piece. By doing so, this manuscript (i) conjectures an issue for keeping sequential order in a network of neurons, or a network made of larger assemblies of neurons in the face of delayed interconnections, and (ii) its possible resolution by generating timestamps through actual movements. To do so introduce this idea, we use a toy model of a ring of three neuronal units, connected by delayed interactions (Figure 1).

**How to get sequential order, or present time, from inside our heads?**

The question of time management within biological systems, and among them by the brain, encompasses multiple functions, processes, and scales (See Golombek et al. 2014; Kelso, 1995; Pittayakanchit et al., 2018; Schöner, 2002; Tallota & Doyère 2020; Winfree, 1980).

Here we will not be concerned with temporal processing, dealing for example with time duration (Safaie et al., 2020). For acting in the world, the physical interaction of the body with the physical environment plays a prominent role (Salinas, 2006). In line with the so-called embodiment approach to perception, action, and cognition (see: Clark, 1998; Kelso, 1995; Port & Van Gelder 1995; Schöner & Spencer, 2016; Turvey et al., 1981; Varela et al., 1991; Warren, 2006), we assume that this coupling to the environment has important consequences on elementary functions, beyond goal directed behavior, or perception-action per se, and underlying sensorimotor functions. The ecological approach is focused on direct perception, following Gibson (1966). It is conjectured that one pervasive organization of the brain prevents autonomy from action in the environment: The presence of coupling delays in the communication between its connected elements. Delays are ubiquitous in the brain, due to the propagation of action potential along axons plus the synaptic delay. The presence of coupling delays has motivated many studies in computational neurosciences, in relation to neuronal synchronization (Campbell, 2007; Ermentrout and Kopell, 1998). What has received far less attention is their consequence for temporal ordering itself. In this opinion piece we will assert that delays could pose hurdles on the path the brain's activity unfolds to maintain a temporal order between internal activities.

**Clarifying the scope: delays, but not compensation**

To prevent confusion, we hasten to put forth that the limitation analyzed here is not concerned with compensating for synaptic or axonal delays, in a perception-decision process for instance. This issue of compensation of time delays in execution and sensory afferences in this context is a classical topic in motor control (see Miall et al., 1993; Miall and Jackson, 2006; Nagy et al., 2023). A school of thought has long proposed the concept of corollary discharge, or efference copy, or of internal models and prediction, as biological solutions to anticipate and compensate for such delays (Von Holtz and Mittelstaedt, 1950; Wurtz, 2018). In the present theoretical analysis and hypothesis, the issue raised is not about compensation of time delays for motor control, but about a consequence of delays at large in brain networks. Here we suggest that it is unclear how the brain can secure the order of events and their synchronization. This is not merely philosophical. An analogous problem arises in distributed computing, where independent processors lack a shared clock (Lamport, 1978). Noteworthy, Leslie Lamport developed a nonphysical solution to this problem of coordinating distributed operating systems, while here we will stress that physical action could afford efficient timestamping. In the next section, we will introduce the concept of delay explicitly with delayed differential equations (DDE). The basic architecture is tree fold: 1) a ring motif connecting 3 units, 2) a movement production system, and 3) a readout device.

**A minimal model: The motif of three nodes arranged on a ring with connecting delays**

In the most standard way delays can be introduced into dynamics is using delayed differential equations (Equation 1):

$$dX/dt = F(X(t), \beta) + G(X(t-\tau), \lambda) \qquad (1)$$

The derivative of the current state is a function, taking parameters $\beta$ and $\lambda$, where F represents the current state $X(t)$ and G the state at a past time t- tau $X(t-\tau)$, tau being the delay.

This equation is shown here only to provide a clear definition of time delays. In full generality and to avoid confusion, for the conjectured effect of time delays analyzed in the paper, F and G could be linear or nonlinear. However, in the case of brain dynamics, nonlinear functions are justified and mostly used so far, whatever the level of analysis taken, from membrane potential of single neurons to neural fields (Amari, 1977; Hodgkin & Huxley, 1952; see for a presentation Deco et al., 2008, or Rabinovich et al., 2006).

In the following we shortly summarize some background about dynamics taking place in delayed systems treated as DDE. Classical textbook on time delayed systems have pointed out that delayed dynamical systems are infinitely dimensional (see a review in Campbell, 2007). The current state derivative is a function of the whole interval between current time (t) and the past time (t-tau). The initial condition is not a single point in the state space, but a history of past value, from (t-tau to t), therefore corresponding to an interval. Time steps being infinitely small, an infinite number of intermediate past states constitute the interval and affect the current state, thus the left-hand side derivative in (1) (see for a presentation Stepan & Kollar, 2000). The numerical integration of such equations is not trivial, as a history of discrete intermediate time steps up to the delay perturbs the dynamics at each current times (Shampine & Thompson, 2001). In the following we won't deal with the study of such dynamical systems as suchs, like studying the impact of delays on stability of synchronization or on bifurcations, leading to oscillations, or multistability, which is beyond the scope of the present paper (see Campbell, 2007 ; Jirsa & Ding, 2004; Roxin et al., 2005; Zheng & Pikovsky, 2019). If noise, ubiquitous in biological systems, is considered, stochastic differential equations (SDE), are extended by incorporating delays to give SDDE. We have yet no clear idea about how noise could change the conjecture proposed in this piece.

Let's consider now a minimal motif: Three nodes arranged in a ring, each coupled to the others with delays (Figure 1 and Equation (2)).

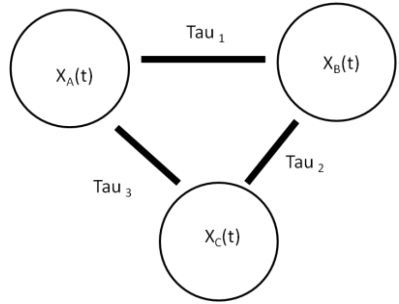

*Figure 1- A cartoon of the architecture of an interconnected system of three neurons, is shown as a ring composed of three nodes A, B, and C, each one with current states $X_i(t)$, i = A, B, C, coupled with time delays tau(j), j = 1, 2, 3. For the present purpose considering the simplest case is sufficient, that is, the nodes and couplings are equivalent, in other words they are invariant by permutation.*

Each node evolves with its own dynamics, steady state, or excitable, or oscillatory for example, which will not be considered in further detail here. Many networks in the brain are of a forward type, and hierarchically organized between an input and an output reached after a series of activities. However, there is considerable experimental evidence and modelling of networks that display spontaneous activity on their own, and don't need an input to become active. A direct illustration of this second type of activity rests on empirical approach around brain oscillations, dating back to the discovery of alpha oscillations, but includes also beta and gamma oscillations, and the continuous efforts to model it (Buzsaki, 2006). A very illustrative case in point here is the seminal model by Wilson and Cowan which can account for self-generated oscillations in two coupled units (Wilson and Cowan, 1972; see for a presentation Destexhe & Sejnowski, 2009). Such a spontaneous oscillatory activity could be considered to generate the activity of each of the nodes X a, b, c in Figure 1. As stated above the dynamics of each node have been mostly modelled so far by nonlinear equations, the coupling between nodes, however, is often chosen to be linear. Our minimal model has the structure of a ring, which highlights the issue of time ordering which is central to the reasoning behind our conjecture. This ring motif is not exotic. Such a motif has been often studied (see for example Bressloff & Coombes, 1999; Collins & Stewart, 1994; Ermentrout 1985; Esnaola-Acebes et al., 2022; Gollo and Breakspear, 2014; Ibrahim et al., 2021; Zheng & Pikovsky, 2019). It appears in central pattern networks (CPG), for which oscillating nodes are required to account for locomotion. However, beyond CPG, as stated above, rings of neurons or of nodes composed of neuronal ensembles are pervasive in the nervous system and widely used in modeling. This scheme of organization, sometimes dubbed a generic motif, can be also generalized beyond nodes representing single neurons or pairs of neurons like in the Wilson and Cowan model, and illustrated by closed-loop connections recruiting distributed areas, linking for instance basal ganglia, cortex and cerebellum, or thalamus and cortical distributed areas (Sporns et al., 1989; Cappe et al., 2009). The description here is kept simple, and the reasoning below is by no way considered as proof but rather as a theoretical proposal, or a conjecture.

**The equations read:**

$$dX_A/dt = F(X_A(t)) + G(X_B(t-tau_1), X_C(t-tau_3))$$

$$dX_B/dt = F(X_B(t)) + G(X_A(t-tau_1), X_C(t-tau_2))$$

$$dX_C/dt = F(X_C(t)) + G(X_A(t-tau_3), X_B(t-tau_2)) \qquad (2)$$

In a symbolic form Equations (2) represent the 3 nodes ring of Figure 1, parameters are dropped for clarity. In this case the couplings are bidirectional or reciprocal, thus B&C act on A, A&C act on B, A&B act on C. The delays, *taus*, can or cannot be equal. It is noteworthy that change from birectional to unidirectional can have drastic effects on the dynamical behavior of rings (Zheng & Pikovsky, 2019), but the behavior of (2) or of possible variants will not be addressed in this paper, our reasoning will be only conceptual and not computational. As far as making units at the nodes more tangible, one may pick up various oscillatory dynamics, for instance based on systems like Stuart-Landau, van der Pol, Wilson-Cowan, or Fitzhugh Nagumo.

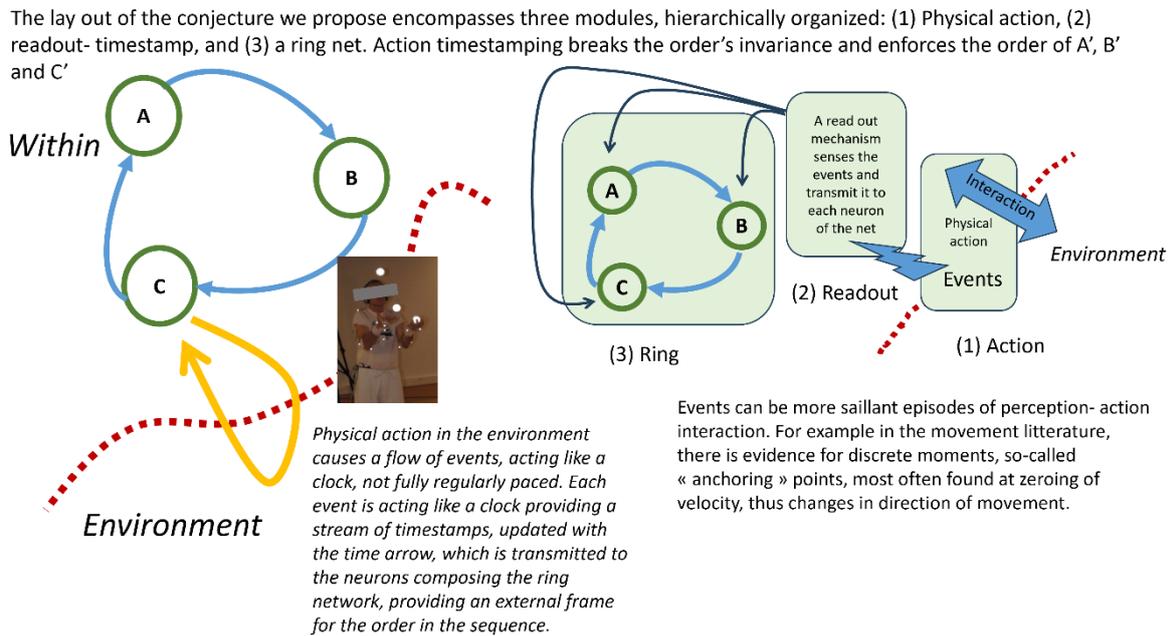

*Figure 2*- *A schematic presenting the general architecture. A ring, a readout, and the timestamping device : an « action » or movement device interacting with the exteranl world. This makes explicit that the ring is dissociated from the timestamping system performing an action.*

**A thought experiment: Timeframe shifting and ambiguity of temporal order**

From our 3 nodes toy model with delays, let's attempt a thought experiment. To emphasize the consequences of time delays in this loop, we assume that this 3 nodes system can be mapped onto another three nodes system with one, say **A** (Xa(t)), functioning in a timeframe that we call *t present* while the 2 others, **B** and **C**, are functioning in past timeframes. Doing so we change the time origin for the different nodes. After applying this transformation, it can be considered that the coupling is no longer delayed but instantaneous (Figure 2). In this case one may imagine that two nodes **B** and **C** are "observed" by the one in the *t present*, that is **A**. **A** could also be considered as marking the time origin, thus relative to it, that is relative to the information **A** could receive, **B** and **C** are functioning in the past. Let's now imagine a sort of time travel experience. Picture yourself as being **A**, you are receiving at each instant information about the state of **B** and **C** coming from their past, like the light of a long dead star reaching the telescope on the earth's surface. However, one could stand also in place of **B** and be receiving information from the past of **A** and **C**. The same applies when standing in the **C** spot on the ring. We won't attempt to define what is present beyond the distinction between past and future (Buonomano and Rovelli, 2021). Put lightly and probably as an expedient, present is not the past anymore and not yet the future. This emphasizes the sequential order notion of time flow, and the causal order between past and present, or present and future. Links to a zoo of themes may be traced out just for the record: Irreversibility (Lynn et al., 2021), Granger-Wiener causality (Granger 1969), non-commutation of rotational movements, including the eyes and the head (see Berger, 2009; Ivancevic, 2004; Tweed et al., 1999), or else delays leading to memory, long range correlations, and non-Markovian processes.

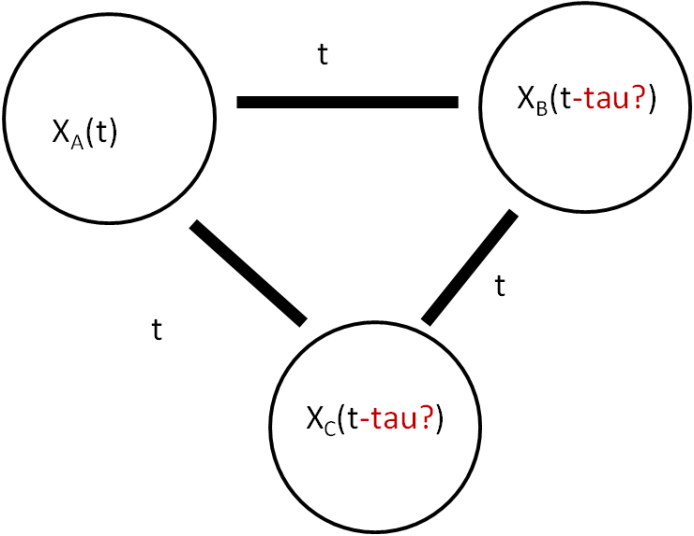

*Figure 3- The 3 nodes ring model after shifting the timeframe origins. In this representation the couplings are instantaneous but each node functions according to distinct timeframe origins. Here the node A($X_A(t)$) is taken as the present. Note that the interrogation along with the delay value tau "tau?" in B and C indicates clearly that the authors are not aware of a proof of equivalence between the two representations of the delayed ring.*

Consider that one can see in this ring model that nodes can be exchanged leaving invariant the network, hence timeframe origin cannot be established in a unique way in the model itself. In such a case, the timeframe ordering is degenerate in the symmetry sense, in other words it is "ambiguous". The consequence is a degeneracy of temporal order. The system contains interactions, causality, and dynamics—but no unique present.

Another way to introduce our hypothesis here is to use a measurement analogy. When several technological measurements devices are used in a scientific or technical set-up, the problem of sensor fusion and synchronization of local clocks has to be solved. This requires a synchronization of several devices, using a trigger, or any type of time stamping procedure. This problem is classical in handling real time and accurate functioning in sensors networks (Nilsson et al., 1998).

The present thought experiment proposes that we have now a system with three entities functioning at distinct times. This emphasizes that if for proper functioning it is required to set what one calls basically the "present" in the ring network, then this has to be done from outside of the ring system. In a nutshell: Present can't be internally obtained. One may think that the interaction with a supplementary network could settle this issue. It is common practice in brain network modeling to set up distinct sub networks, for example to dispatch memory persistence in one network and read out of this memory in another network (See Seung, 1996). We suggest also that this will not solve the problem, and that an initial condition will always be required to get a mark of a timeframe origin. This dependence on the initial state has been identified as a current limit to dynamic, state-dependent networks models of temporal processing (Karmarkar & Buonomano, 2007). A potential parsimonious solution to find grounds for "present" requires to get out, hence, to get contact with the physical environment. We also suggest that there is no other way around, therefore this would be a necessity. If the present argument is solid, the need to tag network dynamics to outside events makes movement in the physical environment mandatory.

It is noteworthy that reference is often made in the present paper to models of so-called temporal processing. We suggest that our conjectural proposition is not restricted to this functional context, the temporal processing in which implicitly or explicitly time has to be dealt with at the level of the function/ behavior. Indeed, the tasks used to address temporal processing are not key in principle here, the effect of delays on the network we focus on is presumably pervasive to large, if not all, brain networks. However, those temporal tasks may be used as practical and natural measurement procedures to investigate which factors perturb or ease the access to $t$ - $present$ in the brain.

**Embodied action as a timestamping device**

We propose that physical action provides the missing reference. We assumed that $t$-$present$ cannot be obtained within interconnected neuron populations, and that this timeframe origin is useful for proper brain functioning. We propose that this $t$-$present$ can be obtained in action-perception loops, that is at each time occurs a relation between movement and its causal perturbation of sensory systems. The recent findings of embodied timing in animal by Safaie et al. (2020) fits in nicely with this framework: Simply put behavior is used to maintain timing (See also Cellini et al., 2015, and Mioni et al., 2016). Let us review in the following some examples of what we dubbed "devices" to get $t$-$present$. In a reaching movement to an object, the physical touch with the object can correspond to the event that provides a $t$-$present$ milestone. The same time milestones are available during upright standing, or even while seated on a chair, as small but permanent motion of the head and torso, or legs, can be sensed via different sensory systems. In the upright postural standing case one can cite of course, among others, the vestibular system which gets access to acceleration due to the gravitational field but also motion consequences in optical flow or acoustic flow can be picked up, or muscle stretching in the torso, hips, and the upper limbs by and large including- legs-ankle-feet. Goal directed and the so-called postural movements can afford this identifiable event, but even simpler devices, like tonic reflex against gravity may be sufficient to signal a contact with the outer world. In summary, every movement produces structured, law-governed perturbations of sensory systems. Touching an object, shifting posture, making a step, catching a ball—each creates an event that is anchored in the physical world and occurs at a definite time.

A) We conjecture that delays create an ambiguity in the order of events in a sequence of variations of activity, in a ring of 3 neurons. This would preclude determining which one's activity is past to whom.

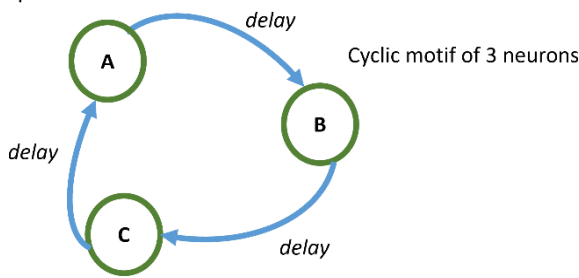

The problem: A', B' and C' are « variations » in activity of neurons A, B and C, respectively. Firing of action potentials is a prime candidate of this activity, which can be oscillatory.

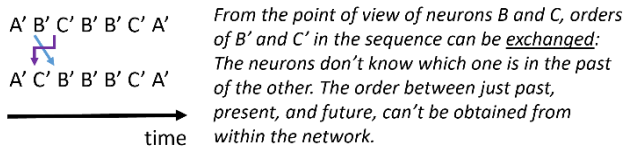

*From the point of view of neurons B and C, orders of B' and C' in the sequence can be exchanged: The neurons don't know which one is in the past of the other. The order between just past, present, and future, can't be obtained from within the network.*

B) Lifting the ambiguity by an action in the outer, « physical », world

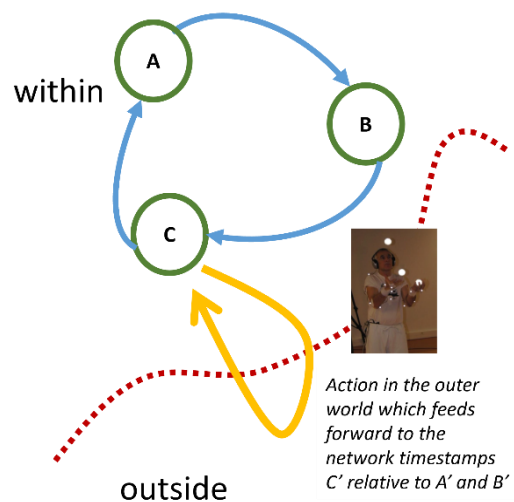

*Action in the outer world which feeds forward to the network timestamps C' relative to A' and B'*

*Figure 4- Left: A schematic view of the core constituent of the conjecture : A) the elementary ring motif with delayed couplings giving rise to the time ordering ambiguity represented her by discrete sequence A', B', C', and Right: A simplified view of the timestamping action (here we took juggling as example) perturbing the ring.*

Examples above provide temporally localized perturbations of the sensory systems by body motion happening at specific times. It may continuously, periodically, or intermittently recur to provide a new initial condition that separates past and future. These perturbations may function as temporal anchors that reset or update the system's temporal reference. A need for recurrence can be predicted as short-term memory suffers known limitations (Ganguli et al., 2008). Limiting cases here are perturbations of the senses which are not self- produced but passive, including any sense, a light, a sound, a tactile contact, a passive limb movement, or an odor. Yet another relevant distinction may concern attention orienting and whether such a passive event is expected or not. A second limiting case could be a touch event self-produced, like touching one hand with the other. At first sight this is not literally a contact with the outer world but a body self-contact.

Furthermore, one can think next to an access to a physical present caused by self-motion not happening at the same time as the current goal-directed action is performed, in an incidental way and concerning a secondary motion of another body part, but instead caused by the goal-directed behavior itself (Datta et al., 2019; Tinbergen, 1963). We can cite in that respect all the synchronization behaviors involving body-world coupling, including rhythmic or discrete cases. Interestingly, the dynamical model of rhythmic synchronization applied to sensorimotor synchronization, that is, a self-sustained rhythmic movement being entrained to an external driving rhythm, is called non autonomous with respect to time. There is a differential equation which is an explicit function of time. That is, time is a variable in the periodic forcing function, for example $F(t) = A \cdot (\cos(\omega \cdot t))$, the dot representing multiplication, that represents the (sensory mediated) perturbation by the external event, on the left-hand side of the dynamical system written in ordinary differential equations.

In a distinct domain one can cite the family of avoidance and control laws (see Schöner & Dose, 1995 or Warren, 2006), in which movement is coupled to an obstacle or to the trajectory of an object. In that respect an interesting candidate device can be to act to meet the time left to avoid falling on the ground; that time limit may set the time landmark we are seeking to get our brain's inner dynamics framed. No matter which coordination pattern is used, which varies between individuals depending on muscle strength, proprioception, body scale and the like, what may count is that a coupling to resolve a physical constraint is set such that it creates an event that again has a date. Hence, a broad class of adjusting posture to avoid falling, or making a step to avoid it, requires attunement, that is a proper adaptation, an adjustment of time scale, to the physics of falling objects, which implies gravitational force and acceleration. To elaborate further, the last example may provide an absolute "metric", or unit, of time duration to the brain. What is the range and average time to fall given your size and height of your center of mass? In the same vein, aiming to catch a falling object requires reference of movement to a physical event in motion independent of the body movement. In the same vein, consider an event caused (or not) by movement which stimulates different sensory systems operating at different speeds and communicating to other parts of the brain with different time delays. For a given event, physiological consequences of sensory perturbations in distinct modalities by physical change are not time aligned, say in a given context the brain faces an average delay of "N" milliseconds between any pair of senses, this may provide also an absolute metric – a unit- for durations.

We briefly reviewed examples above showing that there are several instances for providing our *t-present*. Such encounters with "present" can be regularly updated, estimated, and possibly transiently sustained by short-term memory (Ganguli et al., 2008).

## 3 Discussion

We have suggested that delayed neural interactions introduce an intrinsic ambiguity in temporal ordering that cannot be resolved internally. We suggested above that *acting there* is important for brain functions, and requires recurrent, permanent or intermittent body-world interaction motion events. Is this hypothesis testable? One has to prove that (i) getting *t-present* is important, (ii) some of the devices proposed above are used to get *t-present* at the exclusion of others, and (iii) in the long run getting *t-present* can be related to pathology. Related to the last point one message of radical embodiment, or ecological psychology, is to pay prior attention to the relation of the individual to his environment. In relation to the so-called motor disorders, motor planning and execution are very often poised on the front raw of scientific scrutiny while perception is clearly undermined. Much must be done in that domain attempting instead to take into account the building block of consequences of perception –action coupling, or loops, in another scientific jargon. Disruptions of perception–action coupling may therefore have broad functional consequences, as suggested by work on sensory–motor impairments in Parkinson's disease (Conte et al., 2013).

## 4 Conclusion

The key claim follows directly: a delayed neural network cannot generate a unique present from its internal dynamics alone. One might object that additional subnetworks or readout mechanisms could solve the problem. However, such solutions merely displace the issue. Without external reference, any internal clock still requires an initial condition. Thus, if temporal order matters—and clearly it does—then the reference for "now" must come from outside the network. Maybe the main prediction we come here across is that brain networks involved in or "sensing" this access to present time must be widely connected to many other networks to spread the time origin resetting. Very hypothetically, this could justify large scale integration observed in brains. Second, one may try to perturb the access to the time origin and check the consequences on given functions, temporal ones of course, like a continuation task in simple movement tapping, or self-paced movement, but as suggested every one of them should be impacted. The difficulty lies in perturbing in a way which is complete, that is covering the multiples sources, as we saw time origin devices are manifold.

## References

Amari, S. I. (1977). Dynamics of pattern formation in lateral-inhibition type neural fields. Biological cybernetics, 27, 77-87.

Atay, F. M., & Hutt, A. (2004). Stability and bifurcations in neural fields with finite propagation speed and general connectivity. SIAM Journal on Applied Mathematics, 65, 644-666.

Hodgkin, A. L., & Huxley, A. F. (1952). A quantitative description of membrane current and its application to conduction and excitation in nerve. The Journal of physiology, 117, 500-544.

Badde, S., Ley, P., Rajendran, S. S., Shareef, I., Kekunnaya, R., & Röder, B. (2020). Sensory experience during early sensitive periods shapes cross-modal temporal biases. elife, 9.

Banerjee, A., & Jirsa, V. K. (2007). How do neural connectivity and time delays influence bimanual coordination? Biological cybernetics, 96, 265-278.

Bassett, D., & Sporns, O. (2017). Network neuroscience. Nature Neuroscience, 20, 353–364.

Batista, M., & Peternelj, J. (2006). The Falling Time of an Inverted Plane Pendulum. arXiv preprint physics/0607080.

Betsch, B. Y., Einhäuser, W., Körding, K. P., & König, P. (2004). The world from a cat's perspective– statistics of natural videos. Biological cybernetics, 90, 41-50.

Berger, M. (2009). Géométrie vivante, ou l'échelle de Jacob. Cassini.

Bernstein, N. A. (1967). The Co-ordination and Regulation of Movements Pergamon Press Oxford 1967.

Beuter, A., Bélair, J., Labrie, C. (1993). Feedback and delays in neurological diseases: A modeling study using dynamical systems. Bulletin of mathematical biology, 55, 525-541.

Binetti, N., Siegler, I. A., Bueti, D., Doricchi, F. (2010). Time in motion: Effects of whole-body rotatory accelerations on timekeeping processes. Neuropsychologia, 48, 1842-1852.

Bressloff, P. C., & Coombes, S. (1999). Symmetry and phase-locking in a ring of pulse-coupled oscillators with distributed delays. Physica D: Nonlinear Phenomena, 126, 99-122.

Buonomano, D., & Rovelli, C. (2021). Bridging the neuroscience and physics of time. arXiv [preprint]. doi: 10.48550. arXiv preprint arXiv.2110.01976.

Bullmore, E., & Sporns, O. (2009). Complex brain networks: Graph theoretical analysis of structural and functional systems. Nature reviews neuroscience, 10, 186-198.

Buzsáki, G. (2006). Rhythms of the Brain. Oxford university press.

Campbell, S. A. 2007 Time delays in neural systems. In Handbook of brain connectivity (eds A. R.

McIntosh & V. K. Jirsa), pp. 65–90. Berlin, Germany: Springer.

Capelli, A., Deborne, R., Israël, I. (2007). Temporal intervals production during passive self-motion in darkness. Current psychology letters. Behaviour, brain & cognition, 22.

Cappe, C., Morel, A., Barone, P., Rouiller, E. M. (2009). The thalamocortical projection systems in primate: An anatomical support for multisensory and sensorimotor interplay. Cerebral cortex, 19, 2025-2037.

Chen, Y., Ding, M., Kelso, J. S. (1997). Long memory processes (1/f α type) in human coordination. Physical Review Letters, 79, 4501.

Cellini, N., Mioni, G., Levorato, I., Grondin, S., Stablum, F., & Sarlo, M. (2015). Heart rate variability helps tracking time more accurately. Brain and cognition, 101, 57-63.

Cisek, P., & Kalaska, J. F. (2010). Neural mechanisms for interacting with a world full of action choices. Annual review of neuroscience, 33, 269-298.

Clark, A. (1998). Being there: Putting brain, body, and world together again. MIT press.

Collins, J. J., Stewart, I. (1994). A group-theoretic approach to rings of coupled biological oscillators. Biological cybernetics, 71, 95-103.

Connes, A., & Rovelli, C. (1994). Von Neumann algebra automorphisms and time-thermodynamics relation in generally covariant quantum theories. Classical and Quantum Gravity, 11, 2899.

Conte, A., Khan, N., Defazio, G., Rothwell, J. C., Berardelli, A. (2013). Pathophysiology of somatosensory abnormalities in Parkinson disease. Nature Reviews Neurology, 9, 687.

Dallal, N. L., Yin, B., Nekovářová, T., Stuchlík, A., Meck, W. H. (2015). Impact of vestibular lesions on allocentric navigation and interval timing: the role of self-initiated motion in spatial-temporal integration. Timing & Time Perception, 3, 269-305.

Datta, S. R., Anderson, D. J., Branson, K., Perona, P., & Leifer, A. (2019). Computational neuroethology: a call to action. Neuron, 104, 11-24.

Dayan, P., & Abbott, L. (2001). Theoretical Neuroscience: Computational and Mathematical Modeling of Neural Systems. Cambridge, MA, USA: MIT Press.

Deco, G., Jirsa, V. K., Robinson, P. A., Breakspear, M., & Friston, K. (2008). The dynamic brain: from spiking neurons to neural masses and cortical fields. PLoS computational biology, 4, 1-35.

Destexhe, A., & Sejnowski, T. J. (2009). The Wilson–Cowan model, 36 years later. Biological cybernetics, 101, 1-2.

Dieterich, M., Brandt, T. (2015). The bilateral central vestibular system: Its pathways, functions, and disorders. Ann. N. Y. Acad. Sci., 1343, 10–26.

Dong, D. W., & Atick, J. J. (1995). Statistics of natural time-varying images. Network: computation in neural systems, 6, 345.

Eagleman, D. M., Peter, U. T., Buonomano, D., Janssen, P., Nobre, A. C., Holcombe, A. O. (2005). Time and the brain: how subjective time relates to neural time. Journal of Neuroscience, 25, 10369-10371.

Earl, M. G., Strogatz, S. H. (2003). Synchronization in oscillator networks with delayed coupling: A stability criterion. Physical Review E, 67, 036204.

Edelman, G. M. (1993). Neural Darwinism: selection and reentrant signaling in higher brain function. Neuron, 10, 115-125.

Ermentrout, G. B. (1985). The behavior of rings of coupled oscillators. Journal of mathematical biology, 23, 55-74.

Ermentrout G.B. and N. Kopell, N. (1998). Fine structure of neural spiking and synchronization in the presence of conduction delays, PNAS, 95, pp. 1259–64.

Esnaola-Acebes, J. M., Roxin, A., & Wimmer, K. (2022). Flexible integration of continuous sensory evidence in perceptual estimation tasks. Proceedings of the National Academy of Sciences, 119, e2214441119.

Fitzpatrick, R.C, Day, B.L. (2004). Probing the human vestibular system with galvanic stimulation. J. Appl. Physiol. 96, 2301-16.

Friston, K. (2010). The free-energy principle: a unified brain theory?. Nature reviews neuroscience, 11, 127-138.

Galison, P. (2000). Einstein's clocks: The place of time. Critical Inquiry, 26, 355-389.

Ganguli, S., Huh, D., Sompolinsky, H. (2008). Memory traces in dynamical systems. Proceedings of the National Academy of Sciences, 105, 18970-18975.

Ghasia, F. F., & Angelaki, D. E. (2005). Do motoneurons encode the noncommutativity of ocular rotations? Neuron, 47, 281-293.

Gibson, J. J. (1966). The senses considered as perceptual systems. Boston: Houghton Mifflin.

Gollo, L. L., & Breakspear, M. (2014). The frustrated brain: from dynamics on motifs to communities and networks. Philosophical Transactions of the Royal Society B: Biological Sciences, 369, 20130532.

Golombek, D. A., Bussi, I. L., & Agostino, P. V. (2014). Minutes, days and years: molecular interactions among different scales of biological timing. Philosophical Transactions of the Royal Society B: Biological Sciences, 369, 20120465.

Granger, C. W. J. (1969). Investigating Causal Relations by Econometric Models and Cross-spectral Methods. 16

Grieves, R. M., Jedidi-Ayoub, S., Mishchanchuk, K., Liu, A., Renaudineau, S., & Jeffery, K. J. (2020). The place-cell representation of volumetric space in rats. Nature communications, 11, 1-13.

Haggard, P., Clark, S., Kalogeras, J. (2002) Voluntary action and conscious awareness. Nature Neuroscience, 5, 382–385.

Hoppensteadt, F. C., & Izhikevich, E. M. (2012). Weakly connected neural networks (Vol. 126). Springer Science & Business Media.

Holst, E. V. & Mittelstaedt, H. (1950). The reafference principle. *Naturwissenschaften* **37**, 464–467

Hutt, A., & Frank, T. D. (2005). Critical fluctuations and 1/f α-activity of neural fields involving transmission delays. Acta Phys Pol A, 108, 1021-1040.

Ibrahim, M. M., Kamran, M. A., Mannan, M. M. N., Jung, I. H., & Kim, S. (2021). Lag synchronization of coupled time-delayed FitzHugh–Nagumo neural networks via feedback control. Scientific reports, 11, 3884.

Ivancevic, V. (2004). Symplectic rotational geometry in human biomechanics. SIAM review, 46, 455-474.

Jeffery, K. J., Wilson, J. J., Casali, G., & Hayman, R. M. (2015). Neural encoding of large-scale three-dimensional space—properties and constraints. Frontiers in psychology, 6, 927.

Jirsa, V. K., & Ding, M. (2004). Will a large complex system with time delays be stable?. Physical review letters, 93, 070602.

Jörges, B., López-Moliner, J. (2017). Gravity as a strong prior: implications for perception and action. Frontiers in human neuroscience, 11, 203.

Karmarkar, U. R., Buonomano, D. V. (2007). Timing in the absence of clocks: encoding time in neural network states. Neuron, 53, 427-438.

Kelso, J. S. (1995). Dynamic patterns: The self-organization of brain and behavior. MIT press.

Kelso, J. S., Fuchs, A. (2016). The coordination dynamics of mobile conjugate reinforcement. Biological cybernetics, 110, 41-53.

Krakauer, J. W., Ghazanfar, A. A., Gomez-Marin, A., MacIver, M. A., & Poeppel, D. (2017). Neuroscience needs behavior: correcting a reductionist bias. Neuron, 93, 480-490.

Lamport L. (1978). Time, Clocks and the Ordering of Events in a Distributed System. Communications of the ACM, 21, 558–565.

Lee, D. N. (1976). A theory of visual control of braking based on information about time-to-collision. Perception, 5, 437-459.

Lee, D. N., & Reddish, P. E. (1981). Plummeting gannets: A paradigm of ecological optics. Nature, 293, 293-294.

Longtin, A., & Milton, J. G. (1988). Complex oscillations in the human pupil light reflex with "mixed" and delayed feedback. Mathematical Biosciences, 90, 183-199.

Lynn, C. W., Cornblath, E. J., Papadopoulos, L., Bertolero, M. A., & Bassett, D. S. (2021). Broken detailed balance and entropy production in the human brain. Proceedings of the National Academy of Sciences, 118, e2109889118.

Miall, R. C., Weir, D. J., Wolpert, D. M., & Stein, J. F. (1993). Is the cerebellum a smith predictor?. Journal of motor behavior, 25, 203-216.

Miall, R. C., & Jackson, J. K. (2006). Adaptation to visual feedback delays in manual tracking: evidence against the Smith Predictor model of human visually guided action. Experimental Brain Research, 172, 77-84.

Milton, J. G., Cabrera, J. L., Ohira, T. (2008). Unstable dynamical systems: Delays, noise and control. Europhysics Letters, 83, 48001.

Mioni, G., Labonté, K., Cellini, N., & Grondin, S. (2016). Relationship between daily fluctuations of body temperature and the processing of sub-second intervals. Physiology & Behavior, 164, 220-226.

Moscatelli, A., Lacquaniti, F. (2011). The weight of time: gravitational force enhances discrimination of visual motion duration. Journal of Vision, 11, 5-5.

Nadel, L., & Maurer, A. P. (2020). Recalling lashley and reconsolidating hebb. Hippocampus, 30, 776-793.

Nagy, D. J., Milton, J. G., & Insperger, T. (2023). Controlling stick balancing on a linear track: Delayed state feedback or delay-compensating predictor feedback? Biological Cybernetics, 117, 113-127.

Nilsson, J., Bernhardsson, B., & Wittenmark, B. (1998). Stochastic analysis and control of real-time systems with random time delays. Automatica, 34, 57-64.

Penrose, R. (2005). The road to reality: A complete guide to the laws of the universe. Random house.

Pittayakanchit, W., Lu, Z., Chew, J., Rust, M. J., & Murugan, A. (2018). Biophysical clocks face a trade-off between internal and external noise resistance. Elife, 7, e37624.

Port, R. F., & Van Gelder, T. (Eds.). (1995). Mind as motion: Explorations in the dynamics of cognition. MIT press.

Prigogine, I., & Stengers, I. (1984). Order out of Chaos: Man's New Dialogue with Nature, Boulder, CO. New Science Library.

Pulido, C., & Ryan, T. A. (2021). Synaptic vesicle pools are a major hidden resting metabolic burden of nerve terminals. Science Advances, 7, eabi9027.

Rabinovich, M. I., Varona, P., Selverston, A. I., & Abarbanel, H. D. (2006). Dynamical principles in neuroscience. Reviews of modern physics, 78, 1213.

Roman, I. R., Washburn, A., Large, E. W., Chafe, C., & Fujioka, T. (2019). Delayed feedback embedded in perception-action coordination cycles results in anticipation behavior during synchronized rhythmic action: A dynamical systems approach. PLoS computational biology, 15, e1007371.

Roxin, A., Brunel, N., & Hansel, D. (2005). Role of delays in shaping spatiotemporal dynamics of neuronal activity in large networks. Physical review letters, 94, 238103.

Safaie, M., ..., & Robbe, D. (2020). Turning the body into a clock: Accurate timing is facilitated by simple stereotyped interactions with the environment. Proceedings of the National Academy of Sciences, 117, 13084-13093.

Salinas, E. (2006). How behavioral constraints may determine optimal sensory representations. PLoS biology, 4, e387.

Schöner, G. (2020). The dynamics of neural populations capture the laws of the mind. Topics in Cognitive Science, 12, 1257-1271.

Schöner, G. (2002). Timing, clocks, and dynamical systems. Brain and cognition, 48, 31-51.

Schöner, G., Dose, M., Engels, C. (1995). Dynamics of behavior: Theory and applications for autonomous robot architectures. Robotics and autonomous systems, 16, 213-245.

Schöner, G., & Spencer, J. P. (2016). Dynamic thinking: A primer on dynamic field theory. Oxford University Press.

Schreiber, T. (2000). Measuring information transfer. Physical review letters, 85, 461.

Semjen, A., Leone, G., Lipshits, M. (1998). Motor timing under microgravity. Acta astronautica, 42, 303-321.

Seung, H. S. (1996). How the brain keeps the eyes still. Proceedings of the National Academy of Sciences, 93, 13339-13344.

Shampine, L. F., Thompson, S. (2001). Solving ddes in matlab. Applied Numerical Mathematics, 37, 441-458.

Spence, C., & Squire, S. (2003). Multisensory integration: maintaining the perception of synchrony. Current Biology, 13, R519-R521.

Sporns, O., Gally, J. A., Reeke, G. N., Edelman, G. M. (1989). Reentrant signaling among simulated neuronal groups leads to coherency in their oscillatory activity. Proceedings of the National Academy of Sciences, 86, 7265-7269.

Stepan, G., & Kollar, L. (2000). Balancing with reflex delay. Mathematical and Computer Modelling, 31, 199-205.

Stetson, C., Cui, X., Montague, P.R., Eagleman, D.M. (2005). Illusory reversal of action and effect. Journal of Vision, 5, 769a.

Tallot, L., & Doyère, V. (2020). Neural encoding of time in the animal brain. Neuroscience & Biobehavioral Reviews, 115, 146-163.

Tass, P., Kurths, J., Rosenblum, M. G., Guasti, G., Hefter, H. (1996). Delay-induced transitions in visually guided movements. Physical Review E, 54, R2224.

Thakur, B., Mukherjee, A., Sen, A., Banerjee, A. (2016). A dynamical framework to relate perceptual variability with multisensory information processing. Scientific reports, 6, 31280.

Tinbergen, N. (1963). On aims and methods of ethology. *Zeitschrift für tierpsychologie*, *20*, 410-433.

Turvey, M. T., Shaw, R. E., Reed, E. S., Mace, W. M. (1981). Ecological laws of perceiving and acting: In reply to Fodor and Pylyshyn (1981). Cognition, 9, 237-304.

Tweed, D. B., Haslwanter, T. P., Happe, V., & Fetter, M. (1999). Non-commutativity in the brain. Nature, 399, 261-263

Varela, F. J., Thompson, E., Rosch, E. (1991). The embodied mind: Cognitive science and human experience. MIT press.

Venkadesan, M., Guckenheimer, J., Valero-Cuevas, F. J. (2007). Manipulating the edge of instability. Journal of biomechanics, 40, 1653-1661.

Warren, W. H. (2006). The dynamics of perception and action. Psychological review, 113, 358.

Winfree, A. T. (1980). The geometry of biological time. New York: Springer.

Wurtz, R. H. (2018). Corollary discharge contributions to perceptual continuity across saccades. *Annual review of vision science*, *4*, 215-237.

Yuste, R. (2015). From the neuron doctrine to neural networks. Nature reviews neuroscience, 16(8), 487-497.

Zheng, C., Pikovsky, A. (2019). Stochastic bursting in unidirectionally delay-coupled noisy excitable systems. arXiv preprint arXiv:1902.06915.